\documentclass[aps, pra, twocolumn, secnumarabic, amssymb, superscriptaddress]{revtex4-1}
\usepackage{amsmath}
\usepackage{amssymb}
\usepackage{braket}
\usepackage{mathtools}
\usepackage{bm}
\usepackage {verbatim}

\begin{document}

\title{Coherent electronic-vibrational dynamics in deuterium bromide probed via attosecond transient absorption spectroscopy}

\author{Yuki Kobayashi}
\email{ykoba@berkeley.edu}
\affiliation{Department of Chemistry, University of California, Berkeley, CA 94720, USA}
\author{Kristina F. Chang}
\affiliation{Department of Chemistry, University of California, Berkeley, CA 94720, USA}
\author{Sonia Marggi Poullain}
\affiliation{Department of Chemistry, University of California, Berkeley, CA 94720, USA}
\affiliation{Departamento de Química Física, Facultad de Ciencias Químicas, Universidad Complutense de Madrid, 28040 Madrid, Spain}
\author{Valeriu Scutelnic}
\affiliation{Department of Chemistry, University of California, Berkeley, CA 94720, USA}
\author{Tao Zeng}
\affiliation{Department of Chemistry, York University, Toronto, Ontario M3J1P3, Canada}
\author{Daniel M. Neumark}
\email{dneumark@berkeley.edu}
\affiliation{Department of Chemistry, University of California, Berkeley, CA 94720, USA}
\affiliation{Chemical Sciences Division, Lawrence Berkeley National Laboratory, Berkeley, CA 94720, USA}
\author{Stephen R. Leone}
\email{srl@berkeley.edu}
\affiliation{Department of Chemistry, University of California, Berkeley, CA 94720, USA}
\affiliation{Chemical Sciences Division, Lawrence Berkeley National Laboratory, Berkeley, CA 94720, USA}
\affiliation{Department of Physics, University of California, Berkeley, CA 94720, USA}
\date{\today}

\begin{abstract}
Ultrafast laser excitation can trigger multiplex coherent dynamics in molecules.
Here, we report attosecond transient absorption experiments addressing simultaneous probing of electronic and vibrational dynamics in a prototype molecule, deuterium bromide (DBr), following its strong-field ionization.
Electronic and vibrational coherences in the ionic X$^2\Pi_{3/2}$ and X$^2\Pi_{1/2}$ states are characterized in the Br-$3d$ core-level absorption spectra via quantum beats with 12.6-fs and 19.9-fs periodicities, respectively.
Polarization scans reveal that the phase of the electronic quantum beats depends on the probe direction, experimentally showing that the coherent electronic motion corresponds to the oscillation of the hole density along the ionization-field direction.
The vibrational quantum beats are found to maintain a relatively constant amplitude, whereas the electronic quantum beats exhibit a partial decrease in time.
Quantum wave-packet simulations show that the decoherence effect from the vibrational motion is insignificant because of the parallel relation between the X$^2\Pi_{3/2}$ and X$^2\Pi_{1/2}$ potentials.
A comparison between the DBr and HBr results suggests that rotation motion is responsible for the decoherence since it leads to initial alignment prepared by the strong-field ionization.
\end{abstract}

\maketitle

Ultrafast laser-matter interactions can create coherent superpositions of rotational, vibrational, or electronic states in molecules.
Pure electronic motion in molecules driven by electronic coherence, which is termed charge migration \cite{Woerner17}, can occur even before nuclear motions set in, and spectroscopic observations of such primary processes have been a central topic in attosecond science \cite{Krausz09,Calegari14,Kraus15}.
Potential implications of electronic coherence in photochemistry have been suggested, for example, in selective cleavage of chemical bonds in ionized peptides \cite{Weinkauf95} and efficient charge transfer in light-harvesting antenna \cite{Engel07}.
Theoretical studies have predicted that laser-based control of charge migration is attainable, enabling ultrafast manipulation of the chemical reactivity of photoexcited molecules \cite{Barth06,Loetstedt14,Golubev15}.

Previous attosecond experiments have successfully provided quantitative and angular-resolved information of coherent electronic dynamics in rare-gas atoms \cite{Goulielmakis10,Fleischer11,Fechner14,Kobayashi18}.
There have also been reports on molecular systems \cite{Fleischer11,Kraus13,Timmers19,Ando19,Makhija20}, but the basic questions of how molecular vibrations influence the manifestation of coherent electronic dynamics have yet to be addressed.
Several factors need to be considered, such as the number of participating vibrational modes, relative time scales of electronic and vibrational motions, and displaced equilibrium geometries in various electronic states.
Some theoretical studies predict electronic coherences will survive relatively long (tens of femtoseconds) against vibrational motions \cite{Despre15,Vacher15,Despre18,Jia19}, whereas others show that immediate decoherence will occur in just a few femtoseconds \cite{Jenkins16A,Vacher17,Arnold17}.
Experiments that present quantitative information of multiplex coherences along with a direct comparison to theories can clarify the fundamental mechanisms of electronic-vibrational dynamics in molecules.

Here, we investigate coherent electronic-vibrational dynamics launched in a prototype molecule, deuterium bromide (DBr), using attosecond transient absorption spectroscopy (Fig. 1(a)) \cite{Geneaux19}.
In the experiment, a few-cycle near-infrared (NIR) pulse strong-field ionizes the molecule and initiates coherent electronic-vibrational dynamics.
The attosecond transient absorption spectra at the Br-$3d$ edge probe the ultrafast coherent dynamics with superb state and time resolution, revealing several quantum beats occurring at 0.1-0.3 eV frequencies.
Electronic-structure calculations and wave-packet simulations are performed to construct theoretical core-level absorption spectra, providing  unambiguous confirmation of electronic and vibrational coherences in the ionic X$^2\Pi_{3/2}$ and X$^2\Pi_{1/2}$ states. 
When the polarization direction of the pump and probe pulses is changed from parallel to perpendicular, the phase of the electronic quantum beats shifts by $\pi$, thereby illustrating that the hole density is oscillating between the aligned and anti-aligned directions with respect to the ionization field.
In both polarization measurements, the vibrational quantum beats maintain a relatively constant amplitude, whereas the electronic quantum beats exhibit a partial decrease occurring on a hundred femtosecond time scale.
Quantum wave-packet simulations show that vibrational motion is not responsible for the observed decrease of the electronic quantum beats, in line with the fact that the X$^2\Pi_{3/2}$ and X$^2\Pi_{1/2}$ potentials are nearly identical in their shapes at the Franck-Condon region.
The loss of rotational alignment prepared by the strong-field ionization is suggested as a probable cause, supported by a mass effect found in a  comparison between the DBr and HBr results. 

Experiments are performed with a table-top attosecond transient absorption apparatus described in Ref. \cite{Kobayashi18}. 
A carrier-envelope phase stable femtosecond titanium:sapphire laser system is operated at 790-nm center wavelength, 1.8-mJ pulse energy, and 1-kHz repetition rate.
The laser output is focused into a neon-filled hollow-core fiber for spectral broadening, and a 4-fs NIR pulse is obtained after phase compensation by chirped mirrors and a 2-mm-thick ammonium dihydrogen phosphate plate \cite{Timmers17}.
Part of the NIR beam (100 $\rm{\mu}$J) is picked off by a broadband beam splitter to be used as the pump pulse for strong-field ionization, and the transmitted remainder (200 $\rm{\mu}$J) is used as a driving pulse for high-harmonic generation in argon to produce attosecond extreme-ultraviolet (XUV) pulses.
The pump field intensity estimated from the focus size ($90$ $\mathrm{\mu}$m diam.) is $5\times10^{14}$ W/cm$^2$.
Thin aluminum filters (200-nm thickness) remove residual NIR pulses after transient absorption and high-harmonic generation.
The center photon energy of the XUV spectrum is tuned around 65 eV to address the Br-$3d$ core-level absorption edge \cite{Cummings96,Puttner95,Johnson97}, and the temporal duration of $\sim200$ as was characterized previously for similar XUV spectra with the streaking method \cite{Timmers17}.
A static gas cell (2-mm length) for transient absorption is filled with DBr at a pressure of 5 Torr.
The 99-\% DBr sample was purchased from Sigma-Aldrich and used without further purification.

We first review the potential energy curves of DBr (Fig. 1(b)).
The potentials are calculated with the spin-orbit generalized multiconfigurational quasidegenerate perturbation theory (SO-GMC-QDPT) \cite{Nakano02,Miyajima06,Zeng17,Kobayashi19HBr,Kobayashi19NaI} implemented in the developer version of GAMESS US \cite{Schmidt93}.
See Supplemental Material \cite{smDBr} for computational details.
The neutral ground state of the molecule is X$^1\Sigma_{0^+}$, and its electronic configuration is $[3d^{10}][\sigma^2\pi^4\sigma^{*0}]$.
The two ionic ground states, X$^2\Pi_{3/2}$ and X$^2\Pi_{1/2}$, arise from the $[3d^{10}][\sigma^2\pi^3\sigma^{*0}]$ configuration, and the associated spin-orbit splitting is 0.328 eV \cite{Yencha98}.
In the experiments, a femtosecond pump pulse (red arrow) strong-field ionizes the molecule and launches coherent wave packets on these ionic ground potentials.
After a controlled delay time $t$, an attosecond probe pulse (blue arrow) interrogates the $3d\rightarrow{}\pi$ core-to-valence transitions and encodes the valence dynamics in the characteristic core-level absorption signals. 
The lowest core-excited states arise from the $[3d^9][\sigma^2\pi^4\sigma^{*0}]$ configuration, which splits into five energy levels ($^2\Delta_{5/2}$, $^2\Pi_{3/2}$, $^2\Sigma_{1/2}$, $^2\Delta_{3/2}$, and $^2\Pi_{1/2}$) due to spin-orbit coupling and ligand-field effects \cite{Bancroft86,Kobayashi19HBr}.

\begin{figure}[tb]
\includegraphics[scale=1.0]{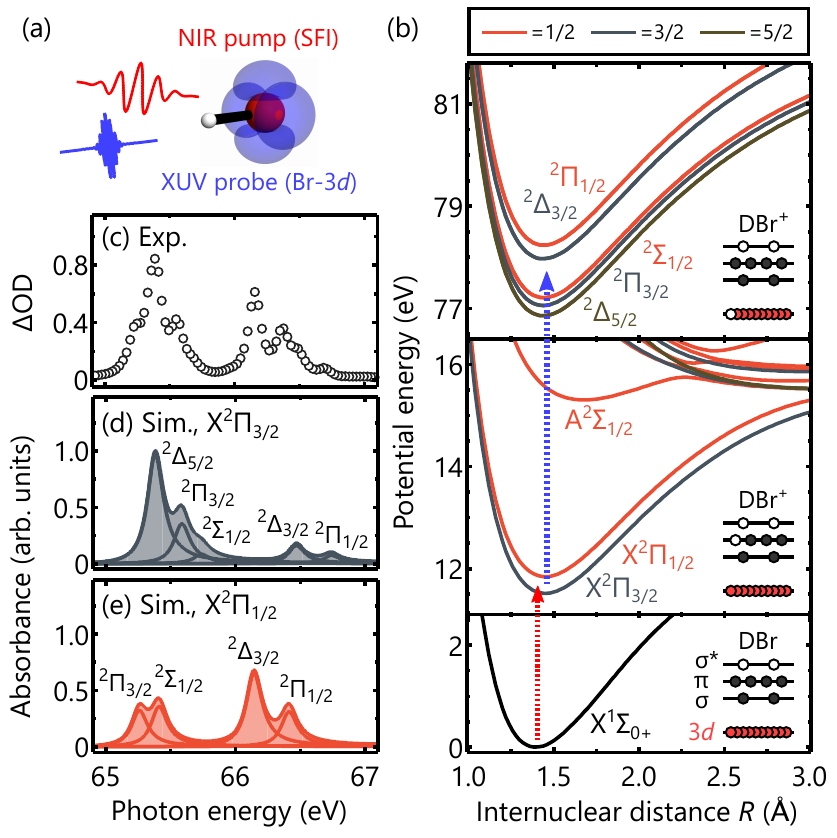}
\caption{\label{fig1}
(a) Pump-probe scheme of the experiment.
A NIR pulse drives strong-field ionization (SFI), and a XUV probe pulse records the dynamics via Br-$3d$ core-level absorption signals.
(b) Potential energy curves of DBr computed for the neutral ground state (bottom), ionic valence states (middle), and ionic core-excited states (top).
Different colors indicate the associated quantum numbers ($\Omega=1/2,3/2,\text{ and }5/2$) of the ionic states.
The red arrows show the ionization step by the strong NIR pulse, and the blue arrow shows the core-to-valence transition by the attosecond XUV pulse.
(c) Experimental transient absorption spectrum of DBr at 200 fs delay time.
(d,e) Simulated absorption signals for the X$^2\Pi_{3/2}$ and X$^2\Pi_{1/2}$ states.
Decomposition into each probe state is denoted.
}
\end{figure}

One of the strengths of core-level transient absorption spectroscopy is its state resolution \cite{Kobayashi19IBr}.
Figure 1(c) shows an experimental transient absorption spectrum of DBr recorded at 200 fs delay time, and Figs. 1(d,e) show the simulated absorption strengths from the X$^2\Pi_{3/2}$ and X$^2\Pi_{1/2}$ states.
A good match is seen between the experiment and simulation, showing the experimental capability to resolve the spin-orbit fine structure of the X$^2\Pi$ states.
Least-squares fitting of the simulated results to the experimental spectrum yields a population distribution of $^2\Pi_{3/2}:{}^2\Pi_{1/2}=0.38\pm0.02:{}0.62\pm0.02$.
Figure 2(a) shows the experimental time-resolved transient absorption spectra.
The measurements are carried out from -10 to 260 fs delay time at intervals of 1.5 fs.
The pump and probe pulses are parallel-polarized in these measurements.
The ionization pump pulse arrives at $t=0$, and the evolution of the ionic dynamics are probed toward positive delays.
Rich oscillation patterns emerge in the entire spectral range, which signify multiple coherent dynamics induced by the strong-field ionization.

\begin{figure}[tb]
\includegraphics[scale=1.0]{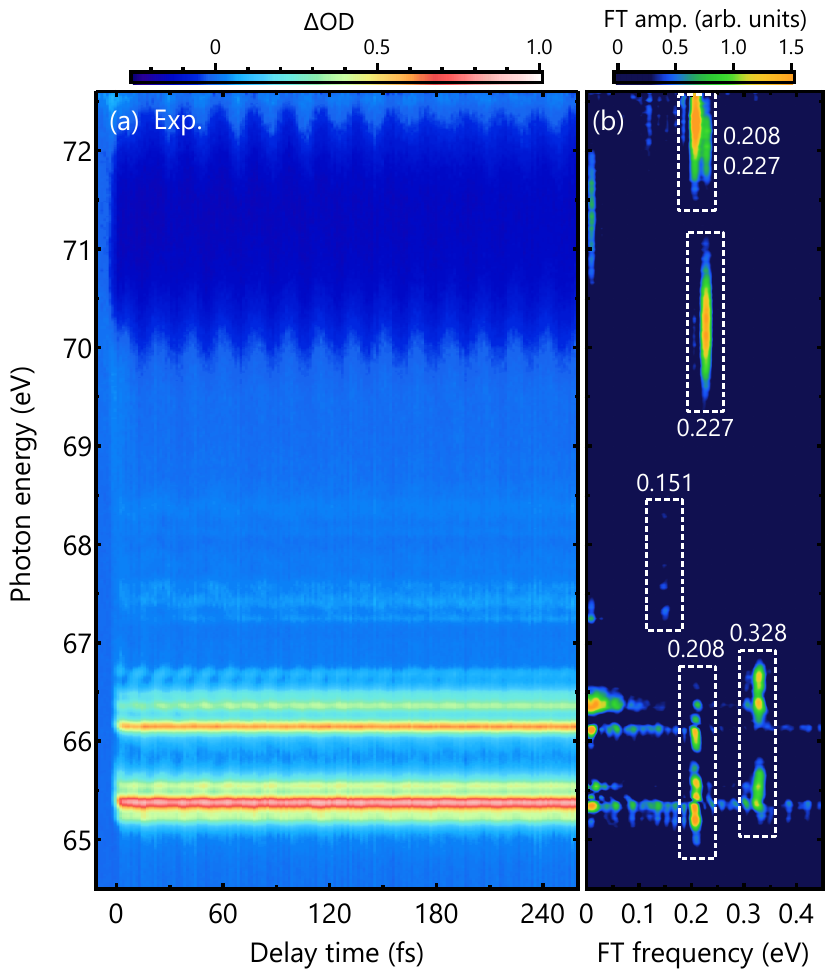}
\caption{\label{fig2}
(a) Experimental delay-dependent transient absorption spectra of DBr.
The pump and probe pulses are polarized to a parallel direction.
Multiple quantum beats are resolved, showing the electronic and vibrational coherences induced by the strong-field ionization.
(b) Fourier transformation (FT) of the experimental spectra along the delay axis.
Main FT components are marked by dashed boxes with the numbers indicating the beat frequencies in units of eV.
}
\end{figure}

A Fourier-transform (FT) analysis is performed along the delay axis to evaluate the beat frequencies, and the results are shown in Fig. 2(b).
In the wide negative depletion from 70 to 72 eV, where the $3d\rightarrow{}\sigma^*$ transition signals emerge  \cite{Kobayashi19HBr}, two beat frequencies of 0.227 and 0.208 eV are observed.
They correspond to the fundamental vibrational frequencies of the neutral X$^1\Sigma$ state (0.234 eV) \cite{Keller54} and the ionic X$^2\Pi$ states (0.209 eV) \cite{Yencha98}, respectively.
The neutral and ionic signals overlap because they both correspond to the $3d\rightarrow{}\sigma^*$ transition and have similar transition energies \cite{Kobayashi19HBr}.
The weak absorption signals from 67 to 68 eV exhibit frequency components of 0.151 eV, and it matches the vibrational frequency of the DBr$^{2+}$ ground state (0.148 eV) \cite{Alagia02,Alagia04}.

Our focus is on the absorption signals of DBr$^+$ that emerge from 65 to 67 eV.
Two beat frequencies are observed for the ionic signals, one at 0.208 eV and the other at 0.328 eV (Fig. 2(b)).
These values match the fundamental vibrational frequency of the X$^2\Pi$ states (0.209 eV) and their spin-orbit splitting (0.328 eV) \cite{Yencha98}, respectively, thus indicating simultaneous vibrational and electronic coherences prepared and probed in the ionized molecule.

In order to corroborate the assignments for the quantum beats, we simulated core-level absorption spectra of the coherent X$^2\Pi_{3/2}$ and X$^2\Pi_{1/2}$ states by numerically solving the time-dependent Schr{\"o}dinger equation for the nuclear motion \cite{smDBr}.
The probe step of core-to-valence transitions is a linear dipole transition, and it can be directly simulated by using the electronic-structure information obtained in the SO-GMC-QDPT calculations. 
The wave-packet simulations further allow one to study the effects of adiabatic vibrational motions on the manifestation of electronic coherence, as will be discussed later.
Figure 3(a) and (b) show the simulated absorption spectra and the Fourier-transform analysis, respectively.
The two frequency components at 0.217 and 0.327 eV match the experimentally resolved quantum beats, providing unambiguous confirmation of their origins as the vibrational and electronic coherences in the ionic X$^2\Pi$ states.
The probing mechanisms of the vibrational and electronic coherences in attosecond transient absorption spectroscopy are illustrated in Figs. 3(c,d).
The vibrational motion translates to the peak shift in the core-level absorption signals \cite{Hosler13}, and the electronic coherence induces constructive or destructive signal variation between the core-to-valence transitions \cite{Goulielmakis10}.
The combined results of experiment and theory establish the powerful ability of attosecond transient absorption spectroscopy to resolve coherent molecular dynamics.

\begin{figure}[tb]
\includegraphics[scale=1.0]{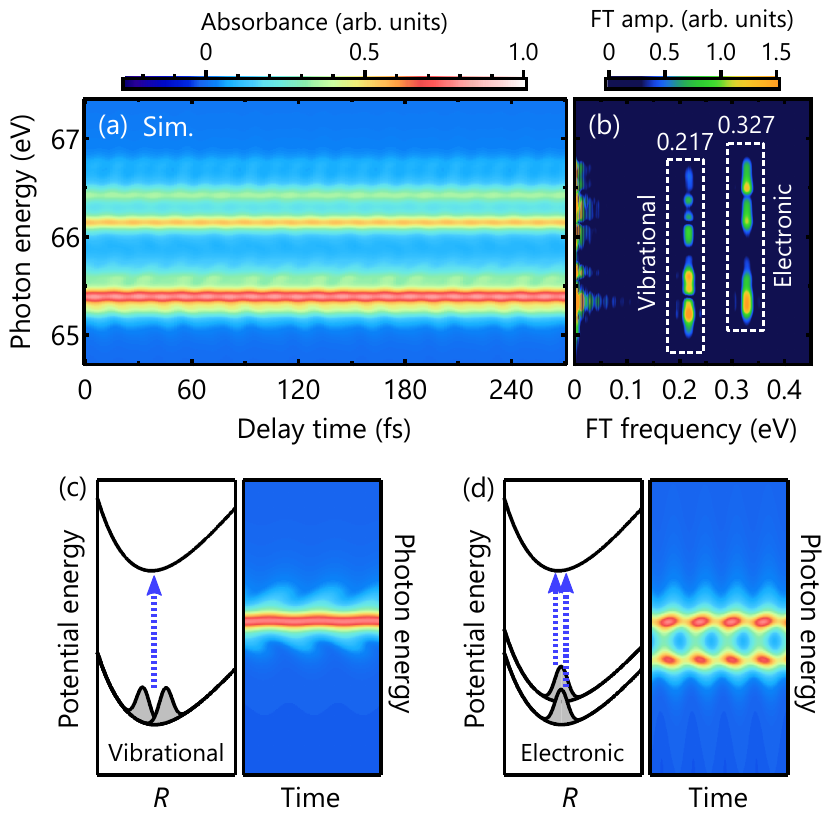}
\caption{\label{fig3}
(a) Simulated delay-dependent transient absorption spectra for the coherent X$^2\Pi_{3/2}$ and X$^2\Pi_{1/2}$ states.
(b) Fourier transformation of the simulated spectra along the delay axis.
The 0.217 eV (vibrational) and 0.327 eV (electronic) components successfully reproduce the experimental quantum beats.
(c,d) Illustration of the probing mechanisms of vibrational and electronic coherences.
}
\end{figure}

A comparison of the FT signal amplitude between the experimental and simulated spectra allows for estimating the degree of electronic coherence, which is defined as $g=|\rho_{ij}|/\sqrt{|\rho_{ii}||\rho_{jj}|}$.
In the equation, $\rho$ is the reduced density matrix of the ionic states, $i$ and $j$ are the state labels, and $g$ is normalized such that $0\leq g \leq 1$ \cite{Santra11}.
The estimated electronic coherence for the present results is $g\approx0.1$ \cite{smDBr}; 
as we will discuss later, vibrational and rotational motion can affect the manifestation of electronic coherence in the absorption spectra, and this value should be taken as a lower limit of the electronic coherence just after the strong-field ionization.


The directionality of the coherent dynamics can further be extracted by changing the probe direction.
Figure 4 shows a comparison between two measurements, in which the pump and probe pulses are polarized in parallel (blue) or perpendicular (yellow) directions with respect to each other.
See Supplemental Material \cite{smDBr} for the full spectra of the perpendicular measurements.
Absorption lineouts are taken at the photon energies representative for the observed quantum beats, either at the centers (electronic, Fig. 3(d)) or edges (vibrational, Fig. 3(c)) of the absorption peaks.
Shown in Fig. 4 are the lineouts at (a) 70.15 eV for the neutral vibrational coherence, (b) 66.69 eV for the ionic electronic coherence, and (c) 65.96 eV for the ionic vibrational coherence.

\begin{figure}[tb]
\includegraphics[scale=1.0]{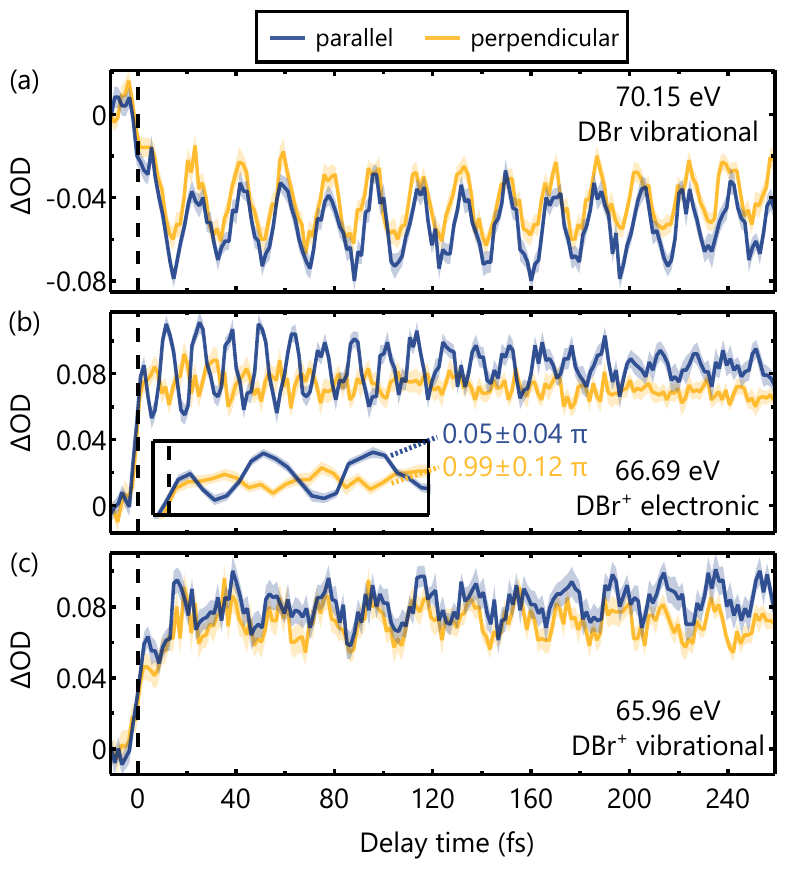}
\caption{\label{fig4}
Comparison between the parallel-polarization results (blue, Fig. 2(a)) and the perpendicular-polarization results (yellow, Fig. S3(a)).
Three lineouts of the absorption signals are taken at (a) 70.15 eV, (b) 66.69 eV, and (c) 65.96 eV.
The shades represent one standard deviation.
The inset in (b) shows an expanded view of the early-time electronic quantum beats.
Quantum beats originating from the vibrational coherences are identical between the two measurements, whereas those originating from the electronic coherence show clear contrast.
}
\end{figure}

The vibrational quantum beats exhibit the same oscillation patterns between the two polarization measurements (Figs. 4(a,c)).
This result is rationalized by the fact that the core-to-valence transition energy is invariant with respect to the probe direction. 
The electronic quantum beats, on the other hand, exhibit a clear variation with polarization (Fig. 4(b)).
Least-squares fitting with a cosine function determines the oscillation phases to be $0.05\pm0.04$ $\pi$ and $0.99\pm0.12$ $\pi$ for the parallel and perpendicular cases, respectively.
The out-of-phase (i.e. $\pi$ phase difference) result qualitatively illustrates that the coherent hole density is switching between aligned and anti-aligned directions with respect to the ionization field. 
Furthermore, the zero initial phase for the parallel case (or $\pi$ initial phase for the perpendicular case) represents that the hole density is most highly aligned along the ionization field direction when the ionization probability is maximized at $t=0$.

Lastly, we address the possible decoherence effects from the molecular vibrations and rotations.
In Figs. 5(a-c), the time evolution of the electronic and vibrational quantum beats is analyzed by taking time-window Fourier transformations of the absorption spectra.
A super-Gaussian function of a 67-fs width was used as a window function, and the FT signals were integrated over the spectral region of the ionic signals (64.95 to 66.88 eV).
Figure 5(d) summarizes the results, showing the integrated and normalized sum of the FT signals for the electronic (orange curves) and vibrational (blue curves) quantum beats.

\begin{figure*}[tb]
\includegraphics[scale=1.0]{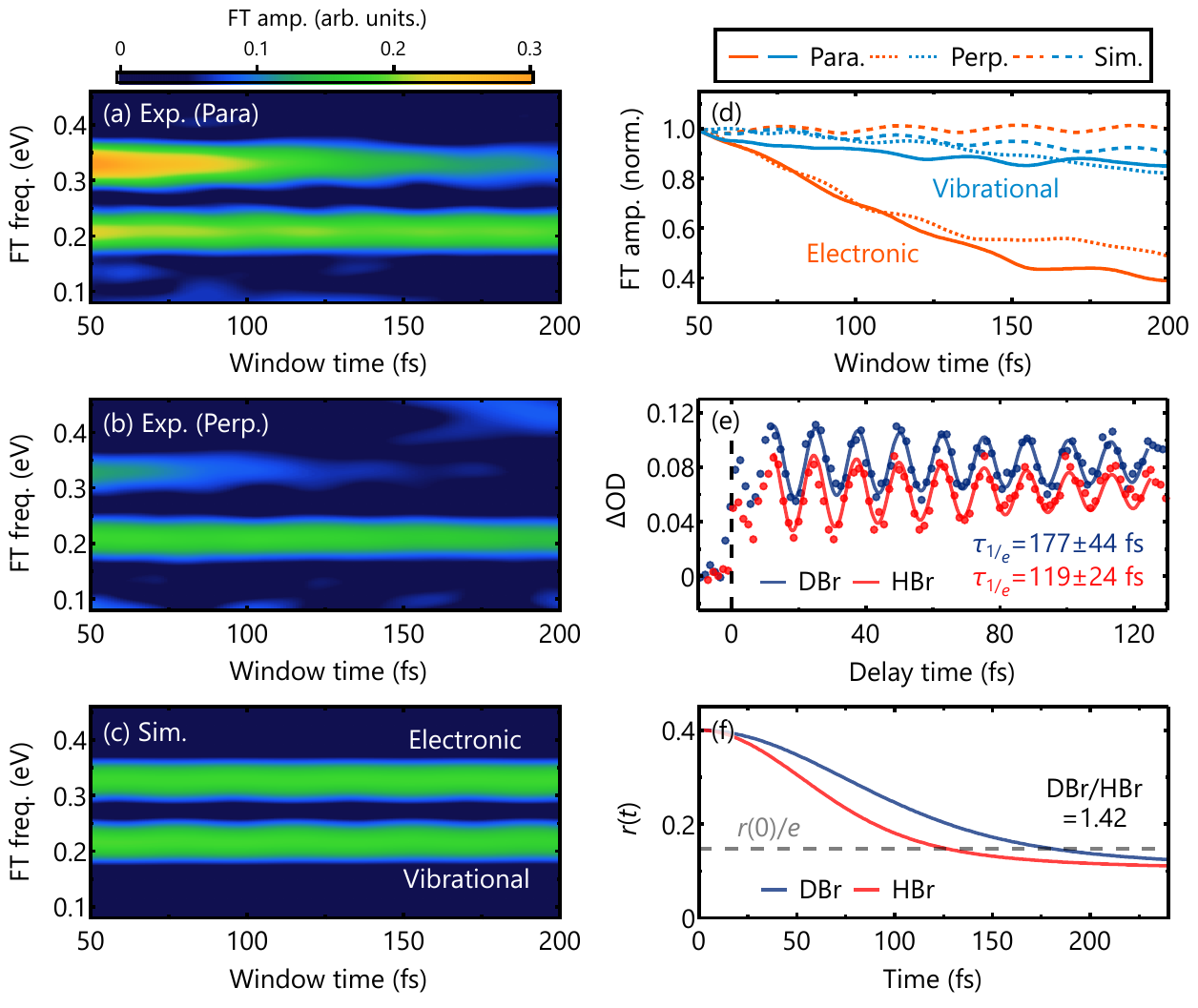}
\caption{\label{fig5}
(a-c) Time-window Fourier transformation analysis of the (a) experimental parallel measurement, (b) experimental perpendicular measurements, and (c) simulated results.
Time evolution of the oscillation amplitude from the electronic and vibrational coherences are extracted.
(d) Normalized signal amplitude for the electronic quantum beats (orange) and vibrational quantum beats (blue).
The electronic quantum beats in the parallel (solid curve) and perpendicular (dotted curve) measurements exhibit a partial decrease, whereas the simulated result (dashed curve) maintains a constant amplitude.
(e) A comparison of the electronic quantum beats between the DBr (blue) and HBr (red) results.
The dots show the experimental absorption signals at 66.7 eV, and the solid curves show the fitting with a cosine function and an exponential decay.
The fitted decay times ($\tau_{1/e}$) are $177\pm44$ fs for DBr, and $119\pm24$ fs for HBr, which yields a ratio of $1.5\pm0.5$.
(f) Time evolution of the anisotropy parameter $r(t)$ calculated for the Boltzmann distributions of DBr and HBr at room temperature.
The decay times are 180 fs for DBr and 127 fs for HBr, and the predicted ratio is 1.42.
The horizontal dashed line shows the value of $r(0)/e$.
}
\end{figure*}

In the experimental results (Figs. 5(a,b)), the vibrational quantum beats maintain a relatively constant amplitude, whereas the amplitude of the electronic quantum beats decreases notably by 100-150 fs.
In the simulated results (Fig. 5(c)), however, the electronic quantum beats maintain a constant amplitude throughout the simulated delay time.
Note that the quantum wave-packet simulations fully take into account the adiabatic vibrational motions.
The contrasting result shows that the vibrational motion is not responsible for the observed decrease in the electronic quantum beats. 
This is explained by the fact that the X$^2\Pi_{3/2}$ and X$^2\Pi_{1/2}$ states share similar potential shapes, with their harmonic frequencies differing only by 5.7 cm$^{-1}$ (1735.5 vs 1729.8 cm$^{-1}$) \cite{Yencha98}, and the spatial overlap or the relative phase between the two wave packets is hardly disturbed by the vibrational motion within the measured delay time \cite{Vacher17}.

The observed partial decrease in the electronic quantum beats is reminiscent of the time evolution of molecular alignment prepared by laser excitation at time zero \cite{Dantus89,Dantus90,Felker92}.
Qualitatively, the hole density can be viewed as initially oscillating between the aligned and anti-aligned directions along the ionization field direction, as revealed in the comparison between the parallel and perpendicular measurements (Fig. 4(b)). 
When the molecular alignment is lost, the hole dynamics will correspond to a breathing motion of spherical density, which will yield smaller absorption variation for the attosecond probe pulse. 

An indirect signature of the rotational effects on the decrease of the electronic quantum beats is found in a comparison between the DBr and HBr measurements (Fig. 5(e)).
See Supplemental Material \cite{smDBr} for the full spectra of the HBr measurements.
A technical issue with HBr is that the vibrational and electronic quantum beats have similar frequencies (0.291 and 0.328 eV, respectively) \cite{Yencha98}, and here the absorption lineouts are taken at 66.7 eV, where only the electronic quantum beats are observed. 
The HBr result exhibits electronic quantum beats at the same frequency and phase as in the DBr result (Fig. 5(e)).
The time scales of the decrease in the quantum beats are analyzed by fitting the experimental signals with a convolution of a cosine function and an exponential decay.
The extracted time constants are $\tau_{1/e}=177\pm44$ fs for DBr and $119\pm24$ fs for HBr, which yields a ratio of $1.5\pm0.5$.
Full quantum-mechanical treatments of electronic-vibrational-rotational dynamics are beyond the scope of this study, and here we provide a comparison to an anisotropy parameter $r(t)$ \cite{Felker86,Dantus89} ($r(t)=0.1$ corresponds to an isotropic distribution), calculated for the Boltzmann distributions of DBr and HBr at room temperature assuming even distributions among the $m_J$ sublevels (Fig. 5(f)).
The calculated decay times, which are defined such that $r(\tau_{1/e})=r(0)/e$, are 180 fs for DBr and 127 fs for HBr, and the predicted ratio is 1.42.
The good match in time ratio between the DBr and HBr results supports that the rotational motion underlies the observed decrease in the electronic quantum beats.





Before concluding, we address two issues regarding the rotational dynamics. 
First, in recent experimental studies where core-level absorption spectroscopy was employed \cite{Saito19,Peng19}, the rotational motion manifested itself as variation in the \emph{absorption amplitude}, while electronic quantum beats were unobserved in those experiments.
In the present experiments, the average absorption amplitude was almost invariant throughout the measured delay time (Fig. 4), and the effect of rotational motion was observed, instead, in the \emph{oscillation amplitude} of the electronic quantum beats (Fig. 4(b)).
These results suggest that even if the hole density is isotropic when averaged in time, the hole-density motion driven by electronic coherence can be polarized and thus serves as a sensitive probe of rotational motion.
Second, an unequivocal evidence of rotational motions would be the observation of alignment revivals \cite{Dantus89,Felker92,Seideman99,RoscaPruna01}.
However, in our auxiliary measurements with HBr, no clear signature of the revival is observed \cite{smDBr}.
Rotational wave packets are usually observed in the neutral ground state of a target molecule, whereas in the present experiments two ionic states and vibrational motions therein are excited along with the possible rotational motions.
These additional complexities may prevent observing the revival features.




In summary, we presented experimental characterization of coherent electronic-vibrational dynamics of DBr$^+$.
The electronic quantum beats are revealed to be unperturbed by the vibrational motion.
This result highlights the importance of inspecting potential differences along the vibrational coordinates, which determine to what extent the spatial overlap and/or the phase relation between the wave packets can be disturbed by the vibrational motions \cite{Vacher17}.
The degree of the electronic coherence in DBr$^+$ is estimated to be $g\approx0.1$; in a previous strong-field ionization experiment on krypton atoms \cite{Goulielmakis10}, a much higher degree of electronic coherence ($g\approx0.6$) was recorded despite its larger coherence bandwidth (0.67 eV).
This contrast shows that the natural spread of nuclear wave packets in molecules hampers the preparation of electronic coherences even for simple diatomic systems. 
The observed decrease in the electronic quantum beats is attributed to the loss of the initial molecular alignment prepared by the strong-field ionization. 
With the ongoing efforts to extend the attosecond spectrum to the water-window regime \cite{Ren18}, we foresee more applications will be made with attosecond transient absorption spectroscopy for coherent electronic-nuclear dynamics in polyatomic systems. 


This work was supported by the US Army Research Office (ARO) (W911NF-14-1-0383) (Y.K., K.F.C., D.M.N, S.R.L.) and the National Science Foundation (NSF) (CHE-1660417) (Y.K., K.F.C., S.R.L.).
Some of the computations by Y.K. were performed using workstations at the Molecular Graphics and Computation Facility (MGCF) at UC Berkeley, which is funded by the National Institutes of Health (NIH) (S10OD023532).
T.Z. acknowledges the Natural Sciences and Engineering Research Council (NSERC) of Canada for research funding (Grant No. RGPIN- 2016-06276) and also York University for the start-up grant (Grant No. 481333).
Y.K. acknowledges support from the Funai Overseas Scholarship.
S.M.P. acknowledges support from the European Union’s Horizon 2020 research and innovation programme under the Marie Sklodowska-Curie grant (842539, ATTO-CONTROL).
V.S. acknowledges support from the Swiss National Science Foundation (P2ELP2\_184414).

\bibliographystyle{apsrev4-1.bst}
%

\end{document}